# POWER LAW SIGNATURE IN INDONESIAN LEGISLATIVE ELECTION 1999-2004

Hokky Situngkir[*]
(hokky@elka.ee.itb.ac.id)
Dept. Computational Sociology
Bandung Fe Institute

**Abstract**
We analyzed the distribution of the result of Indonesian 1999 and 2004 legislative election in order to have the statistical properties or any stylized facts. The paper shows that the analysis results the power-law signature persistently in the election of the House of Representative in 1999 and 2004 and more clearly in the result of Regional Representative Council that held for the first time in 2004. It shows that the socio-political system is a complex system constituted by interacting citizens and accentuates universality that the social evolution dynamically turns to be self-organized in critical state.

*It's not the voting that's democracy; it's the counting.*
Tom Stoppard, Jumpers

**1. Background**
Interval the year 1999 to 2004 is the period of social and political reform in Indonesia. The General Election 1999 is believed to be the first fair election in Indonesia after escaping from 32 years of dictatorship regime. There are 48 political parties joined the election for DPR (House of Representatives) and DPRD I & II (regional and city councils). In 2004, the political reform has urged to change the rule of political system including the election system. The political system has changed to be the bicameralism system while General Election is held in order to let citizens choose directly the members of DPR (House of Representatives), DPRD I & II (city councils) and DPD (the Regional Representative Council). Directly is by means of choose not only from the collection of political parties but choose directly the individuals to be seated in certain political insitutions. It is obvious that the election 2004 become a hope for a better political system in Indonesia.

---

[*] Author is researcher in Bandung Fe Institute, Indonesia. Selected-papers are available in personal homepage: http://www.geocities.com/quicchote

The aim of the paper is to show the changes of statistical properties of the General Election 1999 and 2004 as the voting system changes. We will concern with the legislative election since the direct voting for executive power (presidency) occurs for the first time in 2004, thus we do not have yet any reference to be compared with. In short, the paper will try to answer the question how the socio-political changes as reflected by the statistical properties of the election result happens on the changes of election rule.

As we know from previous works (Situngkir, 2003), political system as we perceived is came out as emergent phenomena from the interacting citizens. Thus, the election result analyzed by using statistics is the facts comes out from interacting political agents vote for aggregate new political system. This is the main idea of social complexity: to look for the pattern of the emergent phenomena from interacting agents. Eventually, we find that the distributional pattern follows the power-law distribution that indicates the character of self-organized in critical condition as the threshold to the chaotic regime (Schuster, 1995:70-72).

We use election result data that accessible from the web-site of General Election Comission (Komisi Pemilihan Umum, 2004). We choose several regional (based on provice) election result, i.e.: Province of Aceh, Province of Sumatera Utara, Province of Sumatera Barat, Province of Sumatera Selatan, Province of Lampung, Province of Jambi, Province of Bengkulu, Province of Riau, Province of Jakarta, Province of Yogyakarta, Province of Nusa Tenggara Barat and Timur as sample of the whole nation-wide election districts. In the other hand, we use the result of the election for DPD (Regional Representative Council) of 2004 election result that taken from all of the election districts. We normalized the votes of each parties and candidates concerning the majority votes of each region and group all of the result of each district in the double-logarithmic plot to see the distribution of votes. We note that we are not dealing with who the winner of the election and the structural political pattern emerges from it, since it will be analyzed in other research paper. What we concern mostly about in the paper is the social analysis as reflected from the data.

## 2. Statistical Model

Let us assume that each candidate has normal distribution of their ability to persuade voters to choose her, denoted by probability $c$; while $p(c)$ is the probability density of persuasion ability equals to $c$.

We can write,

$$p(c) = \frac{1}{\sqrt{2\pi}\sigma} e^{-\frac{(c-\mu)^2}{2\sigma^2}} \qquad (1)$$

where $\mu$ and $\sigma$ are mean and standard deviation respectively of the distribution of $c$. Next, we can say that $f_i(c)$ as pre-election campaign $i$-th process, while then

$$f_1(c) = c_1 \qquad (2)$$

$$f_2(c) = c_2 f_1(c) = c_1 c_2 \qquad (3)$$

and so on up to many pre-election processes, while the average votes gained by candidate in $n$-th subprocess is

$$f_n(c) = c_1 c_2 ... c_n \qquad (4)$$

The total votes gained by candidates are then can be assumed as the multiplicative process since the total votes gained by candidate eventually in the election after large amount of $n$ is

$$v = f_n(c) = c_1 c_2 ... c_n \qquad (5)$$

While $v$ denotes the total votes for candidates in the ballot, and $N(v)$ as the number of candidates that received the fraction votes $v$, we can have the histogram of $N(v)$ vs $v$ as the statistical distribution of the whole voting process. This explanation was also proposed by Filho, et. al. (2002) on finding the power-law characteristic in Brazilian election.

It is very interesting that the histogram resulted ($N(v)$ vs $v$) is not a gaussian one but a power-law characters showing that the election result is somehow a self-organized and critical social conditions. For the election of the member of Regional Representative Council (DPD: Dewan Perwakilan Daerah) that elected for the first time in 2004, the distribution of the number of candidates $N$ receiving fraction of votes $v$ followed a power-law $N(v) \sim v^\alpha$ with fitted $\alpha \approx 1$. This fact can be seen in figure 1.

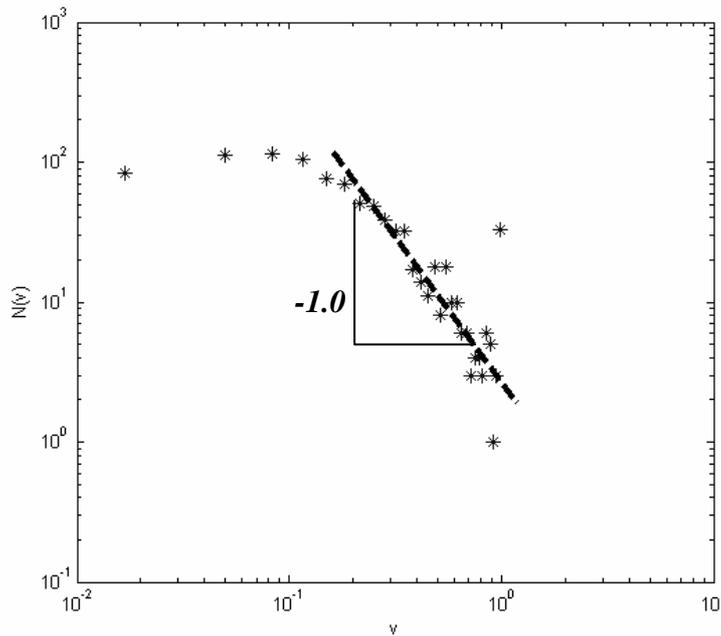

Figure 1
The double-log histogram of the result of DPD (Regional Representative Council) 2004 election.

In the other hand, the result of election for House of Representatives is also power-law but with approximated exponent lower than *1*. The signature of the power-law distribution is not only seen in the result of 2004 election but also in the previous (1999) one. The difference of statistical facts between election 1999 and 2004 is only about other coefficient of the distribution equation. As described in fgure 2, it is clear that the the fitting power distribution of 2004 election is laid above the 1999 one.

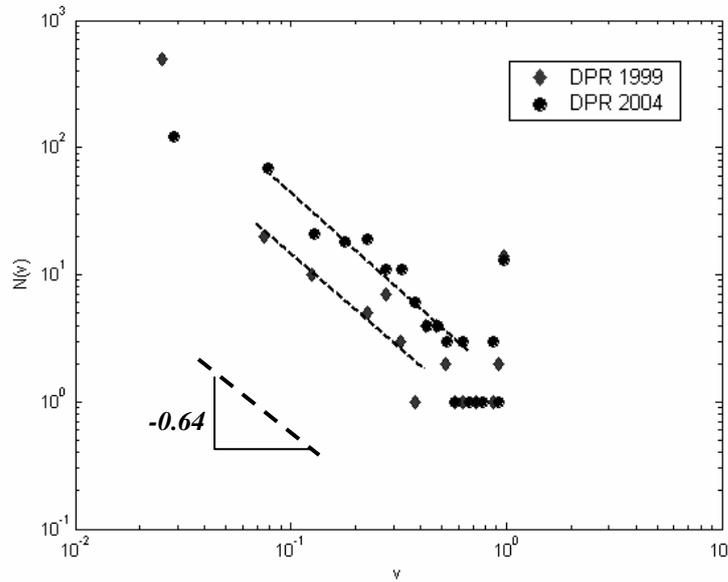

**Figure 2**
The power-law signature in the election result for DPR (House of Representative ) in 1999 and 2004.

## 3. Discussions

In general, the power law distribution can be written as

$$N(v) \approx C v^{-\alpha} \qquad (6)$$

or in double-logarithmic form

$$\log N(v) \approx \log C - \alpha \log v \qquad (7)$$

hence we have a straight line fit in the double-log scales. Statistically, this means that

$$P_r[V = v] \sim v^{-\alpha} \qquad (8)$$

the probability of a candidate gains $v$ votes equals to $v^{-\alpha}$, and in this specific case, the exponent goes around the unity ($\alpha \approx 1$), as described in the previous work (Situngkir & Surya, 2003).

An important lesson we can get from this fact is that social dynamics representing by the the statistical properties of election result are a self-organized criticallity. Every system (including social system) will be persistently evolve towards the critical state. The critical state is signatured by the response functions depict no characteristic scales (Bak, 1997). The power-law distribution is scale-free, means that the character of the critical system does not depend upon time and distance. A sign that showed the robust character of critical state, specifically in social system.

This is somehow interesting and revealed by the result of general election for DPR (the house of representatives) in Indonesia held in 1999 and in 2004. By no concerning about who wins in the elections, we can see that the the exponent coefficients ($\alpha$) in the two respective years are not changed, but the normalization coefficient (figure 2). This shows that the pattern of the election process is just the same in the two different years, despite there are major changes in the rule of voting used in the elections, as being explained above. This is also relating to the result of well-known voting model, e.g.: Snzajd model, that shows how a favorite candidate or party tends to become majority from minority by the behavior of voters to choose candidate or party that mostly chosen in their social network (bahera & Schweitzer, 2003). Free from the economic, socio-political condition and the most popular candidate or party, the pattern tends to be persistent in the critical state. Nonetheless, it is surprising that the changes in the rule of election in 1999 and 2004 does not change the pattern.

**4. Conclusion**
We show that there is power-law pattern in the distribution of number of votes gained in Indonesian election results persistently in the 1999 and 2004 election. This pattern is interpreted as the social robustness in Indonesian political system since the major changes in the election mechanism in 1999 and 2004 are not reflected in the statistical properties of the result. This can also mean that the election as perceived by the citizens is the same from the year 1999 and 2004 despite the changes. Moreover, analytically we can say that the power-law indicates and accentuates of the universality of social evolution that is persistently around the characters of the self-organized in critical state.

**Acknowledgement**
The author thanks Surya Research Inc. for financial support during the research, Ivan Mulianta, Tiktik Dewi Sartika, and Deni Khanafiah for the data gathering and processing assistance. All fault remains the author's.